\documentclass[showpacs,showkeys,amsmath,amssymb,preprint]{revtex4-1}

\usepackage{graphicx}
\usepackage{dcolumn}
\usepackage{tabularx}
\usepackage{color}

\begin{document}
\title{Defect engineering of the electronic transport through cuprous oxide interlayers}

\author{Mohamed M.\ Fadlallah$^{1,2,3}$}\
\author{Ulrich Eckern$^1$}\
\author{Udo Schwingenschl\"ogl$^4$}\email{udo.schwingenschlogl@kaust.edu.sa, +966(0)544700080}\
\affiliation{
$^1$Institut f\"ur Physik, Universit\"at Augsburg, 86135 Augsburg, Germany\\
$^2$Centre for Fundamental Physics, Zewail City of Science and Technology, Giza, Egypt\\
$^3$Physics Department, Faculty of Science, Benha University, Benha, Egypt\\
$^4$KAUST, PSE Division, Thuwal 23955-6900, Kingdom of Saudi Arabia}

\begin{abstract}
The electronic transport through Au--(Cu$_{2}$O)$_n$--Au junctions is investigated using first-principles
calculations and the nonequilibrium Green's function method. The effect of varying the thickness (i.e., $n$)
is studied as well as that of point defects and anion substitution.
For all Cu$_{2}$O thicknesses the conductance is more enhanced by bulk-like (in contrast to
near-interface) defects, with the exception of O vacancies and Cl substitutional defects.
A similar transmission
behavior results from Cu deficiency and N substitution, as well as from Cl substitution and N interstitials
for thick Cu$_{2}$O junctions. In agreement with recent experimental observations, it is found that N and Cl
doping enhances the conductance. A
Frenkel defect, i.e., a superposition of an O interstitial and O substitutional defect, leads to a remarkably
high conductance. From the analysis of the defect formation energies, Cu vacancies are found to be particularly
stable, in agreement with earlier experimental and theoretical work.
\end{abstract}

\maketitle


Low production costs and the right band gap are key factors for applications of p-type
semiconductors. Cuprous oxide (Cu$_{2}$O) fulfils both criteria, making the material
interesting for optoelectronic devices \cite{Zakutayev}, coatings \cite{Nguyen}, sensors \cite{Deng}, etc.
It has a simple cubic structure with lattice constant $a=4.27$ \AA\, 
and a direct band gap at the center of the Brillouin zone, experimentally given by
$2.17$ eV; see Ref.~\citenum{Meyer} for a recent review. From the theoretical point of view,
Hartree-Fock and semi-empirical extended
H\"uckel methods have been used for describing the electronic structure, resulting in
massive overestimations of the band gap \cite{Ruiz,Buljan}. On the other hand,
density functional theory underestimates the gap, giving
0.5 to 0.7 eV within the local density approximation (LDA)
\cite{Buljan,Heinemann}, $0.99$ eV using the LDA+U approach \cite{Heinemann}, and
$1.08$ eV using the self-interaction correction method \cite{Filippetti}. The
Heyd-Scuseria-Ernzerhof (HSE06) hybrid functional gives a band gap of $2.02$ eV,
close to the experimental value \cite{Heinemann}. All
the above mentioned first-principles approaches find the valence band maximum to be dominated
by Cu $3d$ states, and the conduction band minimum by Cu $4s$ states, though
hybridization is important for the material properties \cite{Ruiz,Buljan,Nolan}.

Vacancies and interstitial defects in Cu$_{2}$O have been studied experimentally for
many years, for example, O and Cu vacancies as well as Cu interstitials in
Refs.~\citenum{Yoshimura,Xue,Porat,Aggarwal,Kim,Isseroff}.
The effect of doping with nonmetallic anions at the O site has
been investigated for Cl \cite{Musa} and N \cite{Ish1,Li}; when doping with Si,
Cu vacancy sites are preferably occupied \cite{Ish2,Nemoto}.
Cation doping at the Cu site has been reported for various metals,
including Be \cite{Keeffe}, Mn \cite{Wei}, Co \cite{Kale}, Ni \cite{Kik1,Kik2},
Ag \cite{Tseng}, and Cd \cite{Pap1,Pap2}. As concerns the theoretical methods, the
generalized gradient approximation (GGA) has been employed to investigate the effects
of O and Cu vacancies, anti-sites, and interstitials
\cite{Nolan,Soon}. The GGA+U approach has been used to study
the effects of Mn, Fe, Co, and Ni substitutional doping on the electronic and magnetic properties,
indicating the possibility of long-range ferromagnetism when O and Cu vacancies are
introduced \cite{Sieberer}. GGA calculations also have been reported for doping with
the transition metals Ag, Ni, and Zn \cite{Martinez}. Ag substitutional doping turns
out to reduce only the band gap, whereas Zn doping leads to n-type conduction.

Since the performance of a device usually is dominated by interface properties
\cite{Liu,Zhang}, Cu/Cu$_{2}$O and Au/Cu$_{2}$O interfaces have
been addressed in Refs.~\citenum{Blajiev} and \citenum{Liu}, respectively. In particular, 
In$_{2}$S$_3$/Cu$_{2}$O \cite{Jaya} and ZnO/Cu$_{2}$O \cite{Yang} interfaces have been
found to be promising for photovoltaic applications. 
From the experimental point of view, the electrical resistivities of a Cu$_2$O thin film between
two gold leads using SiO$_2$/Si substrates \cite{Ra}, and of a sulfur-treated n-type Cu$_2$O thin
film with Ag, Cu, Au, or Ni front contacts have been measured \cite{Rb}.
However, very little is known about the {\em theoretical understanding} of electronic
transport properties of heterojunctions involving Cu$_2$O.

In order to provide insight into this question, we study in the following 
prototypical Au--(Cu$_{2}$O)$_n$--Au junctions
including different kinds of defects such as vacancies (O or Cu), substitutional, 
interstitial (N or Cl) and Frenkel defects at different positions in the structures.
We carefully optimize the atomic positions,
in order to obtain physically relevant results. Note that in the main text our focus 
is on the optimized structures; but in most figures we include also
the results for the unrelaxed structures for comparison, including some brief
comments in order to highlight relevant changes due to relaxation.
As is apparent from the results, the optimization of the structures strongly affects 
the electronic structure and transport properties such as the transmission coefficient 
and the conductance. Generally bulk defects are more effective for enhancing transport than
near-interface defects. In order to increase the conductance we suggest that interstitial defects
are more suitable than substitutional ones.

We note from the start that formally, one defect for the $n=4$ junction corresponds to a defect
concentration of 3.8 \%. We also wish to emphasize that the comparison of our model results with
experimental data can only be qualitatitive, since the latter strongly depend on the method of
preparation, as well as the concrete substrate---which we are, at least at present, not able
to incorporate in a quantitative way.

\section*{Results}

Using SIESTA \cite{Sol} as electronic structure package, the transport properties are
calculated by the SMEAGOL \cite{Rocha1,Rocha2,Rungger} code, which employs the nonequilibrium
Green's functions method. Periodic boundary conditions are applied perpendicular to
the [100] transport direction. The Perdew-Burke-Ernzerhof version \cite{Perdew} of the GGA
is utilized
for the exchange-correlation potential. Moreover, the energy cutoff is set to 250 Ry,
and Monkhorst-Pack k-meshes of $15\times 15\times 25$ points for the leads and
$15\times 15\times 1$ points for the transport calculations are used. Norm-conserving
pseudopotentials are employed (fully nonlocal Kleinman-Bylander type \cite{Kleinman}
with double-zeta basis set). 
The pseudopotentials for the Cu, O and Cl atoms include nonlinear core corrections
in the exchange correlation potential \cite{Louie}. The basis set
superposition error is corrected using the counterpoise procedure \cite{Boys}.

A lattice strain of 4 \% is introduced by assuming an
epitaxial interface, and therefore setting the Cu$_{2}$O lattice constant to the Au
value of 4.09 \AA. All structural optimizations (relaxations) are based on the conjugate
gradient method, converging the net force on each atom down to 0.04 eV/\AA. The leads
are modeled by two atomic layers of face centered cubic [001] Au; in addition, four
Au layers on each side of the (Cu$_{2}$O)$_n$ layer are included in the scattering region. 
We study Au--(Cu$_{2}$O)$_{n}$--Au junctions comprising $n=1$ to 4 Cu$_{2}$O unit cells,
which corresponds to 3 to 9 atomic layers of Cu$_{2}$O.

Comparison to previously published theoretical results on bulk Cu$_{2}$O and Au shows
good agreement with our calculations for the bulk materials \cite{Heinemann}. The
well-known fact that GGA underestimates the {\em bulk band gap} indeed is an issue
of debate, but we take here the point of view that, since
the band dispersions are qualitatively correctly described by first-principles
calculations, a correct description of {\em transport through small heterostructures}
can be expected, in particular, concerning the defect and doping dependence.

\begin{figure}[t]
\includegraphics[width=0.80\textwidth]{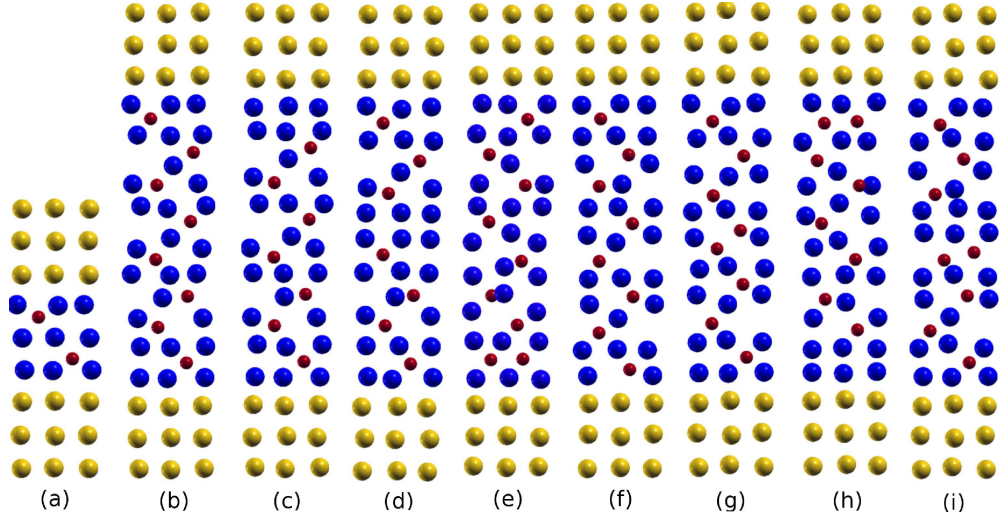}
\caption{(Color online) Structurally optimized Au--(Cu$_{2}$O)$_{n}$--Au heterojunctions
for $n=1$ (pristine, (a)) and $n=4$ (pristine, (b)); O vacancy close to the interface, (c);
O vacancy in the center, (d); interstitial O atom close to the interface, (e); Cu vacancy
close to the interface, (f); Cu vacancy in the center, (g); two configurations of
Frenkel defects, (h), ``near-interface'', and (i), ``bulk-like'').
Yellow, blue, and red spheres represent Au, Cu, and O atoms, respectively.}
\label{fig1}
\end{figure}

The pristine and defective heterojunctions under study are displayed in Fig.\ \ref{fig1}.
We consider, in particular, defects at or near the interface (near-interface), as
well as in the bulk (bulk-like). Concerning O, we study O vacancies near-interface, Fig.\ \ref{fig1}(c),
and bulk-like, Fig.\ \ref{fig1}(d), and O interstitials near-interface, Fig.\ \ref{fig1}(e), and
in the bulk (not presented in Fig.\ \ref{fig1}). Furthermore, Cu vacanies at the interface,
Fig.\ \ref{fig1}(f), and in the bulk, Fig.\ \ref{fig1}(g), are addressed. For Frenkel defects, where
an O atom is removed from a certain layer and inserted into another one, there are two possibilities,
near-interface, Fig.\ \ref{fig1}(h), and bulk-like, Fig.\ \ref{fig1}(i). Finally, N and Cl
substitutional and interstitial defects are also considered at different postions in the structure.

Due to the contact between Au and Cu$_{2}$O, the optimization changes the positions
of the atoms at the interface and in the bulk of the Cu$_{2}$O interlayer; for pristine
junctions, e.g., the Au--Cu distance at the interface is reduced by about 4 \% due
to relaxation. Defects, even those that are two layers away from the interface, alter
the atomic positions further: in fact the additional shift of Cu towards Au at the interface
is maximal for the cases (e), (g), and (h) in Fig.\ \ref{fig1} (another 6 \%).

The defect formation energies give an indication which defects are more likely to be
of experimental relevance than others; it is defined as the energy of the defective structure,
minus the energy of the respective ``constituents'':
\begin{equation}
E_{\rm formation}^{\rm defective} = E_{\rm total}^{\rm defective} - E_{\rm total}^0 , 
\end{equation}
where the subscript ``total'' refers to the respective total energies. For example, for
an X vacancy, $E_{\rm total}^0$ is given by the total energy of the pristine structure,
from which the energy of a single X atom is subtracted:
\begin{equation}
E_{\rm total}^0 = E_{\rm total}^{\rm pristine} - E_{\rm X} \quad {\rm (vacancy)} .
\end{equation}easy 
In particular, we determine $E_{\rm Cu}$ from the Cu bulk energy, while we assume
$E_{\rm O}$ to be half the molecular energy. For a substitutional defect, where the
atom X is replaced by the atom Y, we have
\begin{equation}
E_{\rm total}^0 = E_{\rm total}^{\rm pristine} - E_{\rm X} + E_{\rm Y} \quad {\rm (substitution)} .
\end{equation}
The argumentation for interstitial and Frenkel defects proceeds similarly.

The formation energies---as well as the respective conductances, to be discussed
below---are given in Table I.
It is apparent that O interstitials are by far the most stable
defects among the ones considered, followed by Cu vacancy bulk-like, N interstitial bulk-like,
and bulk-like Frenkel defects. Generally bulk-like defects are more
favorable than interfacial ones except for O vacancies, N interstitial and Cl substitutional defects. 
The stability of structures with Cu vacancies observed here was noticed in previous theoretical 
studies \cite{Raebiger,Scanlon1,Scanlon2}, and was also seen experimentally
\cite{Jongh,Garuthara,Zouaghi,Fujinaka}.

\subsection{Pristine junction}

\begin{figure}[t]
\includegraphics[width=0.48\textwidth]{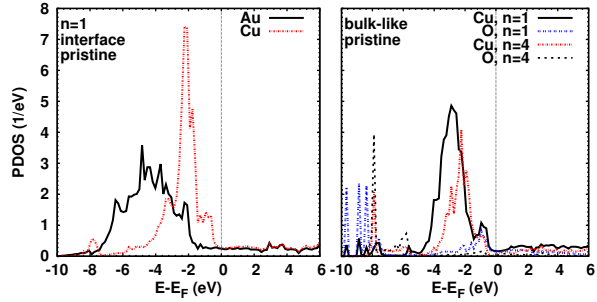}
\caption{(Color online) Projected DOS of atoms at the interface (left) and
in the center of the Cu$_{2}$O region for $n=1$ and $n=4$ (right), after
structural relaxation}
\label{fig2}
\end{figure}

The projected density of states (PDOS) of the interfacial atoms for the 
pristine Au--Cu$_{2}$O--Au heterojunction is shown in Fig.\ \ref{fig2} (left).
Broad Cu $3d$ and Au $5d$ states are apparent, as well as signs of Cu--Au hybridization,
particularly at $-2$ eV.
Figure \ref{fig2} (right) refers to the bulk-like O and Cu atoms in the center of the
Cu$_{2}$O region for $n=1$ and $n=4$. The enhanced Cu--O hybridization around $-1$ eV
is characteristic for the $n=1$ case. Down to about $-5$ eV we observe the expected
dominance of Cu $3d$ states, the DOS structure being broader for $n=1$. 

\begin{figure}[t]
\includegraphics[width=0.48\textwidth]{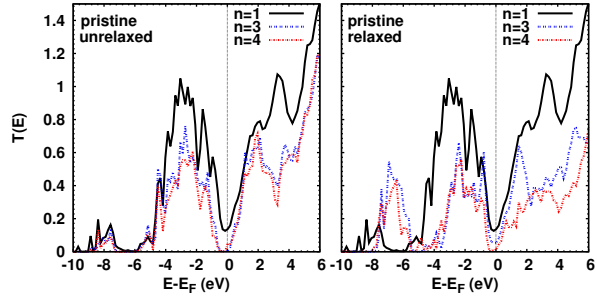}
\caption{(Color online) Transmission coefficient for unrelaxed (left) and relaxed (right)
heterojunctions. While for the unrelaxed system there is hardly any
difference between $n=3$ and 4, the transport gap appears for the relaxed
structure only for $n = 4$. A pronounced transmission around $-7$ eV appears for the
relaxed structure for $n=3$ and 4, due to significant contributions from Au--Cu interface
states (which are absent in the case of $n=1$).
For energies far above $E_F$, the transmisson is somewhat reduced
upon relaxation for $n=3$ and 4. On the other hand, hardly any change due to the
optimization process is observed for $n=1$.}
\label{fig3}
\end{figure}

The transmission coefficients and thickness ($n$) dependencies are shown in
Fig.\ \ref{fig3}. For the optimized structure (right), $T(E)$ reflects the shape
of the DOS below the Fermi energy, $E_F$, compare Fig.\ \ref{fig2}: most notable
are the peak near $-3$ eV (due to the localized Cu levels), as well as the almost
constant DOS above $E_F$ (due to the extended Cu states).  We expect that
the transmission at $E_F$ is dominated by Cu states, since Cu is located next to
Au in the junctions studied. The corresponding conductance, $G=G_{0} \,T(E_{F})$, 
is of the order of $0.15\,G_{0}$ (for $n=1$), where $G_{0}=2e^{2}/h$ denotes the
conductance quantum (the factor 2 is due to the spin). Increasing $n$ to 4, the
conductance drops to $0.014\,G_{0}$.

Above $E_F$, $T(E)$ increases
roughly linearly with energy due to the delocalized Cu 4s states. The
transmission decreases with increasing thickness $n$. A transport gap of
approximately 0.6 eV appears for the relaxed structure when increasing $n$ to 4.
For any thickness studied the heterojunction behaves as an n-type conductor.
In the following, we omit the results for $n=3$ since they are similar to
those for $n=4$.

\subsection{O and Cu defects}

First we discuss the role of O and Cu vacancies and O interstitial defects, cf.\
Fig.\ \ref{fig1}, (c)--(g), as well as of Frenkel defects, which consist of an O vacancy
and an O interstitial, cf.\ Fig.\ \ref{fig1}, (h) and (i).

\begin{figure}[t]
\includegraphics[width=0.48\textwidth]{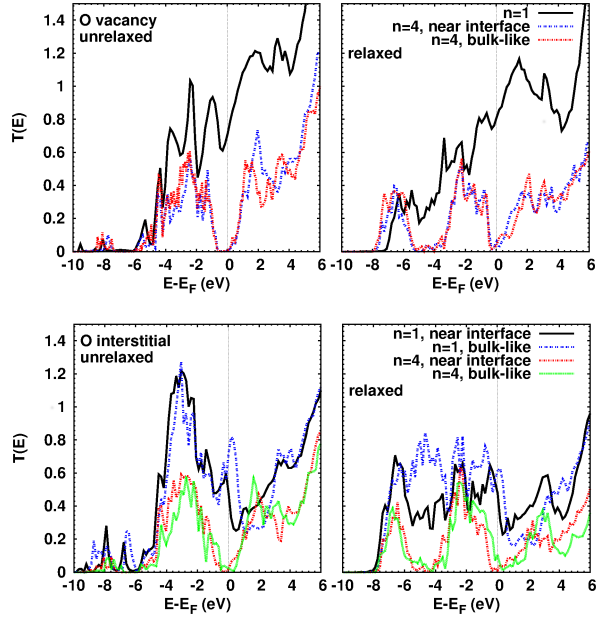}
\caption{(Color online) Transmission coefficient for O vacancies (top) and interstitials
(bottom), before (left) and after (right) structural optimization; compare Fig.\ \ref{fig1},
(c), (d), and (e). The pronounced transport gap found for O vacancies ($n=4$) before relaxation
is filled due to structural optimization. On the other hand, there is hardly any relaxation
effect for vacancies for $n=1$, while it appears largest for O interstitial doping.
The transmission at $E_F$ generally is higher when the atomic positions are
optimized, and that bulk-like defects lead to a higher transmission than near-interface defects.}
\label{fig4}
\end{figure}

Figure \ref{fig4} shows $T(E)$ for O vacancies (top part of the figure)
and O interstitial defects (bottom part of the figure). In the following, we concentrate our
discussion on the optimized structures (right). A major change of the transmission,
compared to the pristine junction, is observable for both cases for $n=1$, namely a shift of
weight from the peak around $-3$ eV towards $E_F$, and an enhancement below $-5$ eV.
However, $T(E)$ is more evenly distributed for an interstitial defect. In addition,
$T(E)$ is enhanced at the Fermi energy by vacancies due to contributions from Au and Cu
states, as is apparent from the respective PDOS (not shown here).
The conductance is $0.82\,G_{0}$, and thus considerably larger than in the pristine case.

For the thicker Cu$_{2}$O layer, $n=4$, we consider O vacancies at the interface
and in the center of the Cu$_{2}$O region, see Fig.\ \ref{fig1}, (c) and (d).
The exact position of the vacancy plays only a minor role for $T(E)$. As compared
to the pristine junction, the transport gap at $E_F$ has closed, but
the conductance is rather small, about $0.063\,G_{0}$ and $0.035\,G_{0}$ for the
near-interface and bulk-like vacancy, respectively.

An O interstitial atom in a thin ($n=1$) Cu$_{2}$O interlayer 
results in a significant $T(E)$ down to $-8$ eV, see Fig.\ \ref{fig4} (rightmost).
Furthermore, the minimum of $T(E)$ around $E_F$ is completely washed out, resulting
in enhanced conductances of $0.41\,G_{0}$ (near-interface) and $0.62\,G_{0}$ (bulk-like).
However, the enhancement is slightly less than what is observed for O vacancies.
For the thick ($n=4$) Cu$_{2}$O layer the effect of an interstitial O atom strongly
depends on its position: the conductance is considerably higher for the bulk-like
than for the near-interface interstitial ($0.11\,G_{0}$ vs.\ $0.02\,G_{0}$).
This fact can be related to a strong contribution of p orbitals, which lead to a
clearly visible peak in the O (bulk-like) PDOS at $E_F$, even larger than the Cu
(bulk-like) PDOS. For a near-interface interstitial, on the other hand, this peak is
located near $-1$ eV. Overall $T(E)$ behaves similar to the pristine junction.

\begin{figure}[t]
\includegraphics[width=0.48\textwidth]{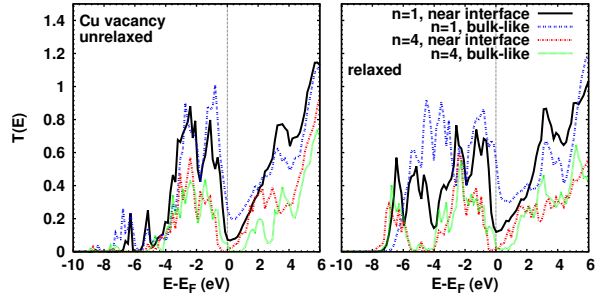}
\caption{(Color online) Transmission coefficients for Cu vacancies, before (left)
and after (right) structural optimization; compare Fig.\ \ref{fig1}, (f) and (g).
While the overall behavior is similar
before and after relaxation, one notes an increase around $-7$ eV and a decrease for
energies far above $E_F$ due to optimization. At $E_F$, the transmission is
higher for $n=1$ compared to $n=4$ and for a bulk-like compared to near-interface
defects. Slightly above $E_F$, a Cu bulk vacancy creates for $n=4$ a strong transport
gap, for both the relaxed and unrelaxed situation.}
\label{fig5}
\end{figure}

In Fig.\ \ref{fig5} we show the transmission coefficients for Cu vacancies; we focus again
the discussion on the optimized structures (right). As compared to the pristine junction,
vacancies enhance $T(E)$ below $-5$ eV for $n=1$. For an interfacial Cu vacancy the change
in $T(E)$ below $E_F$ can be related to the sensitivity of the $d$ bands to disorder,
and the Au($5d$)--Cu($3d$) orbital interaction at the interface \cite{Autes}. The conductance 
is reduced (compared to the pristine junction) to $0.12\,G_{0}$ ($n=1$). For a bulk-like
Cu vacancy an enhanced $T(E)$ is visible between $-5$ and $-4$ eV due to contributions
from Cu 3d bulk states, and the conductance is given by $0.52\,G_{0}$.
An increase of the conductance due to 
Cu vacancies is in agreement with previous work on thin films \cite{Isseroff}.
Independent of its position, a Cu vacancy shifts the minimum of $T(E)$ around
$E_F$ to higher energy.

On the other hand, for a thick ($n=4$) Cu$_{2}$O layer, the gross shape of $T(E)$ for
Cu vacancies is very similar to the case of O vacancies. The conductance,
however, is higher for the bulk-like than for the interfacial Cu vacancy ($0.032\,G_{0}$
vs.\ $0.026\,G_{0}$). Note that the conduction type can be modified by Cu vacancies.

\begin{figure}[t]
\includegraphics[width=0.48\textwidth]{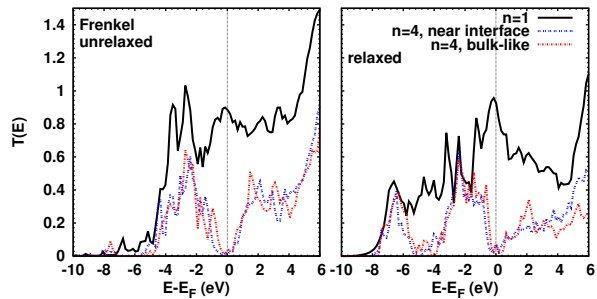}
\caption{(Color online) Transmission coefficients for $n=1$ and 4 heterojunctions
with Frenkel defects, before (left) and after (right) structural optimization. The labels
``near-interface'' and ``bulk-like'' refer to (h) and (i) in Fig.\ \ref{fig1}, respectively.
For $n=1$, of course, there is only one type of Frenkel defect. Due to the optimization,
there is a drastic change in the transmission in the energy range $-6$ eV to $E_F$. A pronounced
peak near $-7$ eV appears for all thicknesses. Again an increase of the
transmission at $E_F$ is found for $n=1$ and 4 due to relaxation, while the transmission above
$E_F$ is reduced.}
\label{fig6}
\end{figure}

Turning to Fig.\ \ref{fig6}, we observe that a Frenkel
defect modifies $T(E)$ substantially for $n=1$, leading to a very high conductance of 
$0.92\,G_{0}$. For $n=4$ we consider the two Frenkel defect configurations
shown in Fig.\ \ref{fig1}, (h) and (i). We find no significant difference in $T(E)$, and obtain
essentially the same conductance (0.070 vs.\ $0.075\,G_{0}$) for both of them. Roughly speaking,
$T(E)$ can be considered a combination of the effects of an O vacancy and an O interstitial.

\subsection{N and Cl doping}

\begin{figure}[t]
\includegraphics[width=0.48\textwidth]{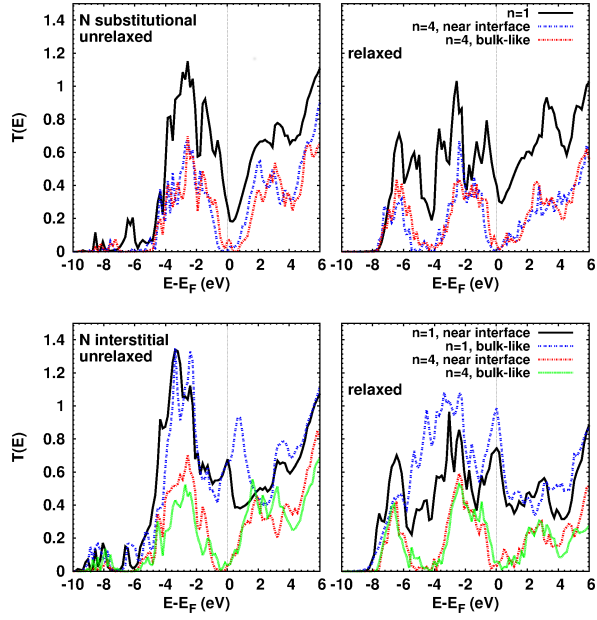}
\caption{(Color online) Transmission coefficients for N substitutional (top) and
interstitial doping (bottom), before (left) and after (right) structural optimization.
For the $n=1$ heterojunction, besides the appearance of strong transmission around
$-7$ eV as discussed before, the shift of the the extrema near the Fermi energy upon
relaxation is remarkable, in particular, for N interstitial doping. The transmission
is found to depend significantly on the location of the N interstitial only for the
optimized case. For $n=4$, it
appears that transmission is slightly higher after relaxation for bulk-like doping
(substitutional or interstitial) compared to near-interface doping.}
\label{fig7}
\end{figure}

The effect of N substitutional and interstitial doping is addressed in Fig.\ \ref{fig7}.
Clearly the onset of $T(E)$ is related to the energetic positions of the N relative to
the Cu states, and the high $T(E)$ around $-6.5$ eV is due to a strong overlap between 
them. Minima in $T(E)$, for example near $-4$ eV, coincide with minima in the
Cu PDOS (not shown here). The shift of the minimum of $T(E)$ around $E_F$
to higher energy is similar to the effect of a Cu vacancy. Conductance 
values of $0.40\,G_{0}$, $0.74\,G_{0}$, and $0.99\,G_{0}$ are obtained for N substitution,
near-interface N interstitial, and bulk-like N interstitial, respectively ($n=1$). Substitution
creates an excess hole, while an interstitial defect creates excess electrons. This explains
why the conductance is higher for the interstitial than for the substitutional case. For
bulk-like N interstitial there are more transmission channels than for the near-interface
N interstitial at $E_F$.

Figure \ref{fig7} also deals with N doping at different positions for $n=4$. Comparing
substitutional and interstitial doping, the overall energy dependence is quite similar
(and similar to the pristine junction), except for subtle differences near $E_F$. As
compared to the pristine junction, the transport gap is slightly reduced, and the
conductance thus a little enhanced, similarly for bulk-like and near-interface substitution, to
about $0.015\,G_0$. Turning to N interstitial defects ($n=4$),
Fig.\ \ref{fig7} shows that the bulk-like position of the defect results in a slightly
higher conductance than the near-interface position, $0.061\,G_0$ vs.\ $0.031\,G_0$.

\begin{figure}[t]
\includegraphics[width=0.48\textwidth]{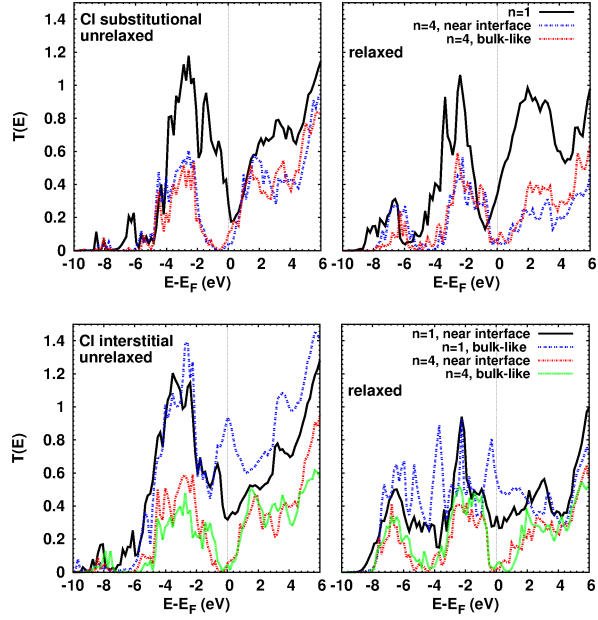}
\caption{(Color online) Transmission coefficients for Cl substitutional (top) and
interstitial doping (bottom), before (left) and after (right) structural optimization.
For substitutional doping, the position of the extrema at $E_F$ for the $n=1$ unrelaxed system
are similar to the N case, compare Fig.\ \ref{fig7}, whereas the situation is different for
$n=4$. For interstitial doping, on the other hand, for both $n=1$ and 4 the behavior of the
unrelaxed stystems is close to the N interstitial. The conductances of the optimized structures
are less or equal to the corresponding unoptimized ones for $n=1$. Again for $n=4$ the conductances
are found to increase upon optimization.}
\label{fig8}
\end{figure}

Figure \ref{fig8} addresses Cl doping. For $n=1$ substitutional doping,
we find close similarities to the pristine junction though the minimum near
$E_F$ has shifted to lower energy, which is opposite to what is found for N doping.
The conductance of about $0.34\,G_{0}$ accordingly is higher than in the pristine junction,
but lower than in the case of N doping. The shift of the minimum of $T(E)$ to lower
energy appears also for the Cl interstitial defect. A bulk-like Cl interstitial
defect leads to a higher conductance ($0.56\,G_0$) than a near-interface Cl interstitial
defect ($0.32\,G_0$).

Interestingly, see Fig.\ \ref{fig8}, for $n=4$ we find again a shift of the minimum
of $T(E)$ near $E_F$ to higher energy (compared to the pristine case)
for the bulk-like Cl substitutional defect. The conductances are given by $0.065\,G_0$
and $0.043\,G_0$ for near-interface and bulk-like substitution, respectively, which
is considerably higher than for N substitution. In contrast to the other cases considered,
the Cl bulk-like substitution has a {\em lower} conductance than the near-interafce one.
The overall results for Cl interstitials are similar to those for N interstitial defects,
but with slightly smaller conductances ($0.022\,G_0$ for the near-interface, and
$0.033\,G_0$ for the bulk-like case). The calculated conductances for $n=4$ are summarized
in Table I.

\section*{Discussion}

Based on the density functional theory and the nonequilibrium Green's function approach, 
a comprehensive study of the electronic structure and transport properties of
Au--(Cu$_{2}$O)$_n$--Au heterojunctions of different thickness, with different point
defects, and with different kinds of anion doping of Cu$_{2}$O has been carried out. As
to be expected, the transmission decreases with inceasing $n$.
Compared to the pristine junction, we find for thin interlayers, $n=1$, for O vacancies
and for interstitial O, N, and Cl defects a drastic change of $T(E)$ around the Fermi energy.
Bulk-like defects are found
to enhance the transport more effectively than near-interface defects---which appears reasonable
since the system has a better chance to adjust to a bulk-like compared to a near-interface
disturbance, thereby reducing scattering. Accordingly, we also find that for all cases considered
the formation energy is lower for bulk-like than for interface or near-interface disturbances.

Experimentally it has been found that N and Cl doping enhances the conductance \cite{Musa,Ish1},
and that interstitial N defects are more effective than N substitution \cite{Li}.
While this 
agrees with our findings, further work is needed to establish the connection between the 
present model studies and the actual experimental situation, where, in particular, the
film thicknesses are much larger than the ones in our model.
Note that for Cl we observe the opposite trend: interstitial defects are less effective than
substitution. In conclusion, our investigation provides indications on how to improve
the electrical and photovoltaic properties of Cu$_{2}$O contacted by two gold leads, namely by
appropriate, preferably bulk-like defect engineering. 
Of course, it must be kept in mind that
the detailed experimental conditions may not always be properly reflected by the 
idealized theoretical modelling.

\acknowledgments{
We acknowledge financial support by the Deutsche Forschungsgemeinschaft (through TRR 80).
Research reported in this publication was supported by the King Abdullah University of
Science and Technology (KAUST).
}

\section*{Author contributions}
M.M.F.\ performed the calculations. All authors contributed to the analysis of the results and the writing
of the manuscript.

\section*{Additional information}
The authors declare no competing financial interests.

\newpage

\begin{table*}[htb]
\caption{Formation energies $\Delta E \equiv E_{\rm formation}^{\rm defective}$ (in units of eV),
and conductances $G$ (in units of $10^{-2}G_0$, $G_0 = 2e^2 /h$, for $n=4$. For comparison, the
conductance of the pristine junction is given by $1.4\times 10^{-2}G_0$ ($n=4$): 
thus the conductance
of the defective junctions can be up to a factor of about 8 (for O interstitials, bulk-like) larger
than for the ``clean'' case.}
\begin{center}
\begin{tabular}{ |c|c|c||c|c|c||c|c|c||c|c|c| } 
\hline
  & $\Delta E$& G & O & $\Delta E$&  G & N & $\Delta E$ &  G & Cl & $\Delta E$ & G \\
\hline
O vacancy,& &            & interst., & &               & interst., & &              & interst.,  & & \\
near interf. & 0.3 & 5.8 & near interf. & $-$0.2 & 2.0 & near interf. & 0.1 & 3.1   & near interf. & $-$0.1 & 2.2 \\
\hline
O vacancy, & &         &interst., & &                  & interst., & &             & interst., & &\\
bulk-like  & 0.4 & 3.2 & bulk-like & $-$1.7 & 11       & bulk-like &  $-$1.1 & 6.1 & bulk-like  & $-$0.2 & 3.3 \\
\hline
Cu vacancy, & &            &Frenkel, & &               & subst., & &              & subst., & &\\
at interf.  & $-$0.3 & 2.6 & near interf. & 0.8 & 7.0  & near interf. & 2.0 & 1.5 & near interf. & 1.1 & 6.5 \\
\hline
Cu vacancy, & &           &Frenkel,  & &               & subst., & &              & subst., &  & \\
bulk-like  & $-$1.6 & 3.2 & bulk-like& $-$0.8 & 7.5    & bulk-like & 1.0 & 1.4    & bulk-like & 1.9 & 4.3 \\
\hline
\end{tabular}
\end{center}
\end{table*}

\newpage

\end{document}